# The effective surface roughness scaling of the gelation surface pattern formation

T. MIZOUE, M. TOKITA, H. HONJO, H. J. BARRAZA, AND H. KATSURAGI

*T. Mizoue, H. Honjo, H. Katsuragi,*

*Department of Applied Science for Electronics and Materials, Kyushu University, Kasuga, Fukuoka 816-8580, Japan*

katsurag@asem.kyushu-u.ac.jp

*M.Tokita,*

*Department of Physics, Kyushu University, Ropponmatsu, Fukuoka 810-8560, Japan*

*H.J. Barraza*

*Unilever R&D, Quarry Road East, Bebington CH63 9HW, United Kingdom*

The surface pattern formation on a gelation surface is analyzed using an effective surface roughness. The spontaneous surface deformation on DiMethylAcrylAmide (DMAA) gelation surface is controlled by temperature, initiator concentration, and ambient oxygen. The effective surface roughness is defined using 2-dimensional photo data to characterize the surface deformation. Parameter dependence of the effective surface roughness is systematically investigated. We find that decrease of ambient oxygen, increase of initiator concentration, and high temperature tend to suppress the surface deformation in almost similar manner. That trend allows us to collapse all the data to a unified master curve. As a result, we finally obtain an empirical scaling form of the effective surface roughness. This scaling is useful to control the degree of surface patterning. However, the actual dynamics of this pattern formation is not still uncovered.

## Introduction

Pattern formation of soft materials is one of the most ubiquitous phenomena in nature [1]. Our body is certainly an example of soft pattern formation. And many dissipative structures appear in soft materials. However, we don't know any general method to characterize/analyze these fascinating pattern formation phenomena properly. Natural patterns have very wide range of diversity. The pattern formation of gel is a typical example of such soft pattern formation.

Since Tanaka et al., have discovered the beautiful surface pattern formation on polymer gels during volume phase transition [2,3], it has been extensively studied both by experiments [4,5] and theories [5,6,8]. In this pattern formation, mechanical instability due to the abrupt volume change is crucial to understand it. Obviously, complex polymer network resulting in intriguing viscoelastic behavior of polymer gel is a main reason of this pattern formation.

On the other hand, recent studies have shown a novel kind of pattern formation which appears during polymer gelation. The one entity is 1-dimesional (1D) pattern formation. Narita and Tokita have found a Liesegang pattern formation on 1D κ-karageenan gel [9]. The diffusion of potassium chloride in κ-karageenan solution is essential process in this pattern formation. The other entity is a quasi-2-dimesional (2D) one. The Acrylamide (AA) gelation on a Petri-dish has shown a spontaneous surface deformation [10]. The competition between the positive feedback of radical polymerization and the inhibition by oxygen is thought a main reason of this pattern formation. This surface deformation is actually 3-dimenstional (3D) phenomena, while the gelation occurs in quasi 2D space, i.e., the situation is not very simple. The observed pattern looks like wrinkles on brains or surface pattern on reptiles. This similarity reminds us that the nature might be using this kind of spontaneous surface deformation. The study of this pattern formation is thus important both by means of gelation dynamics itself, and pattern formation dynamics in bio soft matter. In Ref.[10], the reaction diffusion dynamics is presented to understand this pattern formation. However, there remains some difficulties to explain that pattern formation using a usual reaction diffusion dynamics. More detail experiments and universality check are necessary to model this pattern formation correctly.

In this paper, we will focus on this quasi 2D pattern formation with radical polymerization. First, we check the universality of this pattern formation using some kinds of monomer that undergo the radical polymerization gelation. In addition, we will characterize the degree of surface patterning using the effective surface roughness (ESR) of 2D picture. The experimental conditions are systematically varied and resultant surface deformation is characterized by the ESR. Finally, we find an empirical unified scaling of the ESR.

## Experimental

The experimental system is simple. Pre-gel solution is poured onto a Petri-dish, and it is left about 2 hours. Then, spontaneous surface deformation occurs depending on the experimental condition. If Ref.[10], only AA gel is used as a monomer. Thus, we try some monomers that are able to undergo radical polymerization. Concretely, Sodium acrylate (SA, $M_w = 94.05$), N-Isopropylacrylamide (NIPA, $M_w = 113.16$), and Dimethylacrylamide (DMAA, $M_w = 99.13$) are used as monomers. In all cases, Methylenbisacrylamide (BIS) constitutes cross-link. And, Ammonium persulfate (APS) and Tetramethylethlyenediamine (TEMD) are used as an initiator, and an accelerator of the radical polymerization, respectively.

We mainly controlled the concentration of initiator, temperature, and ambient oxygen. Sample preparation and temperature control method are same as Ref.[10]. Here, we additionally control the ambient oxygen concentration using an airtight chamber and $O_2$, $N_2$ gas cylinders. After 2 hours polymerization, resulting surface patterns are taken by a CCD camera, and the photos are processed by a PC.

In Fig. 1, typical patterns observed with each monomer are shown. It is hard to see the clear surface deformation with SA and NIPA gel. They seem to have very weak surface deformation instability. This is due to that the lower critical solution temperature (LCST) in NIPA is close to the experimental temperature (30 degree Celsius), and the remaining inhibitor in commercial SA. The NIPA gel is very sensitive to temperature. The detail temperature dependence of NIPA gel pattern formation is open for future problem. Since the DMAA doesn't have such difficulties, it shows relatively clear surface pattern. The observed pattern is more or less similar to the AA one. Thus, we decide to study the DMAA surface deformation pattern formation, in this study.

## Phase Diagram

As a next step, we systematically make DMAA gel slabs under various experimental conditions and compose the phase diagram as shown in Fig. 2. The specific experimental conditions are shown in Table 1. Qualitative structure of this phase diagram is very similar to the AA case [10]. The surface deformation pattern appears between the completely flat gelation ("Flat") and the incomplete gelation ("Not-gelation"). This means that the inhibition of polymerization is an crucial process to make surface instability. In addition, the large scale buckling can be observed in the marginal region between the "Surface deformation" and "Not-gelation". In the surface deformation pattern, bottom plane of the gel slab is flat (i.e., the deformation is limited on the top surface), while the buckling includes bottom deformation. The origin of this large scale buckling has not been clarified yet, since it is more difficult than the surface deformation pattern. Here, we focus on surface deformation pattern again, because we don't know the details of this even easier case. Noticeable feature of Fig. 2 phase diagram is wider patterning region in relatively low temperature regime. It is a characteristic feature of DMAA pattern formation different from the AA. Moreover, the clear stripe patterns cannot be observed in DMAA surface pattern formation.

## Effective surface roughness analysis

In order to quantify the degree of surface deformation, we employ the standard deviation of 2D photos. We can recognize the surface deformation through the contrast of 2D photos (like Fig. 3). This suggests that the standard deviation of 2D photos can be used as an indicator of the surface deformation degree. In Fig. 3, typical 2D pictures with varying initiator concentration are presented. As can be seen in Fig. 3, increasing initiator concentration tends to suppress the surface deformation. Moreover, buckling can be observed in very low initiator cases. We don't use such buckling regime. To characterize these photos, central part (1,000 pix. * 1,000 pix.) of raw data (3,072 pix. * 2,304 pix.) is extracted for each photo. Then, the data are translated to 8 bit gray scale, and finally the standard deviation and average of the photo intensity values are computed. We define this standard deviation the effective surface roughness (ESR).

First, we vary the initiator concentration and temperature under atmospheric condition (ambient oxygen concentration is about 21 %). Since the surface deformation regime is limited as shown in Fig. 2 phase diagram, the completely independent change of initiator concentration and temperature is difficult. We have to adjust both of them simultaneously to create surface deformation pattern. We show the computed average and ESR values in Fig. 4. While almost the constant average intensity is confirmed in Fig. 4(a), increasing ESR is observed for decreasing initiator concentration (Fig. 4(b)). This trend is consistent with pictures in Fig. 3. The almost constant average indicates the reproducible lighting and/or other external noise factors. The negative correlation between the ESR and initiator concentration implies that the more the initiator, the more stable the polymerization. As a result, a uniform flat slab is created in the case with sufficient amount of initiator.

Next, the ambient oxygen and temperature are maintained to create surface deformed slabs. We have to vary the initiator concentration as well to create clear surface deformation, owing to the narrow patterning regime (same reason as previous Fig. 4 case). The measured average intensity and ESR are shown in Fig. 5. Constant average intensity is the same trend as Fig. 4 case. However, the ESR and oxygen concentration shows positive correlation. This trend is consistent with the inhibition effect of oxygen in radical polymerization. The oxygen scavenges and stops the radical polymerization, so that the flat surface is inhomogeneous and unstable. This is presumably the principal origin of surface instability.

This oxygen inhibitor effect corresponds to the counter against the initiator stabilizing effect. And, the concave structure of ESR is similar in Figs. 4 and 5. From these factors, one can expect that all ESR curves can be unified to a single master curve. To unify all ESR data, here we use a cubic function as a fitting form,

$$[ESR] = \alpha + \left( \frac{1}{\sigma_T} \cdot \frac{[O_2]}{[I]} \right)^3 \quad (1)$$

where $\sigma_T$ is a parameter depending on the temperature $T$ (degree Celesius), and $\alpha$ denotes the background noise level from camera, lighting, etc. The $[I]$ and $[O_2]$ correspond to initiator concentration (g/l) and ambient oxygen concentration (%), respectively. The reason we choose the cubic function is that the exponential form is too fast and the quadratic form is too slow to fit and unify the ESR data. In principle, the exponent of the fitting function can be assumed as a free fitting parameter. However, that would not improve the data collapse so well, in spite of adding one more free fitting parameter. One of the reasons to use the cubic function is to reduce free fitting parameters. The form of Eq. (1) is a completely empirical form. It is found in a heuristic way as mentioned above.

Anyway, we can fit the all ESR data to Eq. (1) and obtained the following scaling,

$$\frac{1}{\sigma_T} = 0.38 T^{-0.41}. \quad (2)$$

The actual relation between $\sigma_T$ and $T$ is shown in Fig. 6(a). The data scatters a little, but they agree with a trend described by Eq. (2) (gray curve in Fig. 6(a)). The scaling form Eq. (2) means that the following scaling variable $f([O_2],[I],T)$ is useful to unify the data,

$$f([O_2],[I],T) \sim \frac{[O_2]}{[I] \cdot T^{0.41}}. \quad (3)$$

The final data collapse using this scaling variable is displayed in Fig. 6(b). The data show a little scattering again, but the scaling function captures the trend of the ESR data. The concrete function form of the final scaling (gray curve in Fig. 6(b)) is as following,

$$[ESR] = 3.7 = 6.6\left(\frac{[O_2]}{[I]\cdot T^{0.41}}\right)^3 . \qquad (4)$$

This scaling form enables us to approximately predict the degree of surface deformation, while it is only an empirical law. The relation between this empirical scaling form and physical mechanism of surface deformation is a future problem.

## Discussion

There are theoretical studies for the gel surface deformation [6,7,8,11]. Most of them were focusing on the volume phase transition. However, the present surface deformation clearly occurs at the gelation (polymerization) stage. In that stage, polymer blobs diffusion and polymerization reaction are basic process. When the polymerization is inhibited at the gel-solution interface by oxygen, the swelling of polymer network might cause deformation locally. Thus, the diffusion, reaction, and network swelling, all factors might have to be coupled to explain this surface deformation pattern formation. As long as we see the gelation process by eyes, the gelation seems to grow from the bottom to the surface. More direct observation of gelation surface is necessary to understand the pattern formation properly.

In this study, we use the cubic function to fit the ESR data. The cubic function is concave, and it means that abrupt growth of ESR occurs at a certain $f([O_2],[I],T)$. Such concave growth inevitably leads the divergent ESR. Then the large scale buckling happens to avoid the divergence. In fact, the gel slab with $f([O_2],[I],T) > 25$ tends to show buckling. The detail investigation of buckling instability is also an open problem. Besides, the characteristic length scale study like Ref. [10] might be helpful to discuss the pattern formation dynamics.

Self-organizing pattern formation is a frontier in material science. Most of all self-organized patterns show nano- or micro- meter order structures [12,13]. Such micro structures are of course practical to design the functional materials. Contrastively, the pattern we report here has macro (millimeter order) structure. Although the benefit of such macro structure is still unclear in terms of functionality, it is visible by naked eye and easy to control. We believe that macro structures in soft matter also has a great potential for the application.

## Conclusions

We systematically performed the experiments on surface deformation occurring on free gelation surface. We define the ESR (Effective Surface Roughness) of the gel slabs utilizing 2D digital photo data, in a very simple way. We varied the initiator concentration, ambient oxygen concentration, and temperature as control parameters of gelation, and measured the ESR for each slab. As a result, we obtained the simple empirical scaling form Eq. (4) to characterize the degree of surface deformation. While we can estimate the surface deformation amount owing to this empirical scaling, the detail relation between this form and surface deformation dynamics is still unsolved.

Table 1, Experimental conditions

| | |
|---|---|
| DMAA [ml] | 1.8 |
| BIS [mg] | 4 |
| TEMD [μl] | 70 |
| water [ml] | 11 |
| APS [mg] | 0-60 |
| Temperature [C] | 10 - 60 |

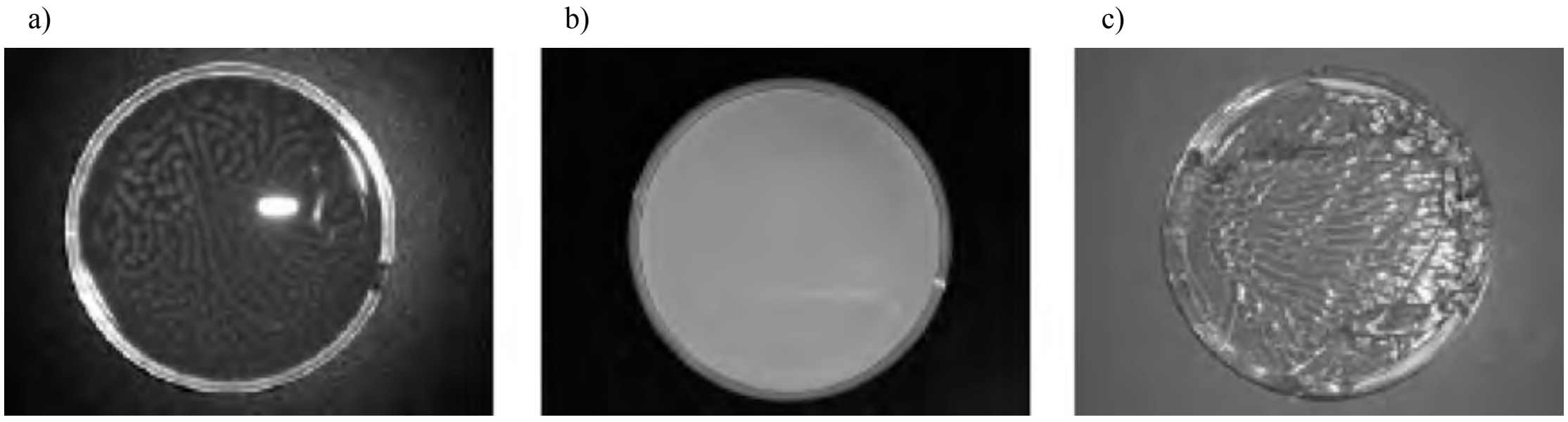

Fig. 1 Gel patterns with **a** SA (=2 mg), **b** NIPA (=1.9 mg), and **c** DMAA (=2.4 mg) monomers. In all cases, 6 mg BIS, 70 μl TEMD, and 10 mg APS are dissolved to 12 ml deionized water under the room temperature. SA and NIPA gel show very marginal patterns, while DMAA gel shows clear pattern.

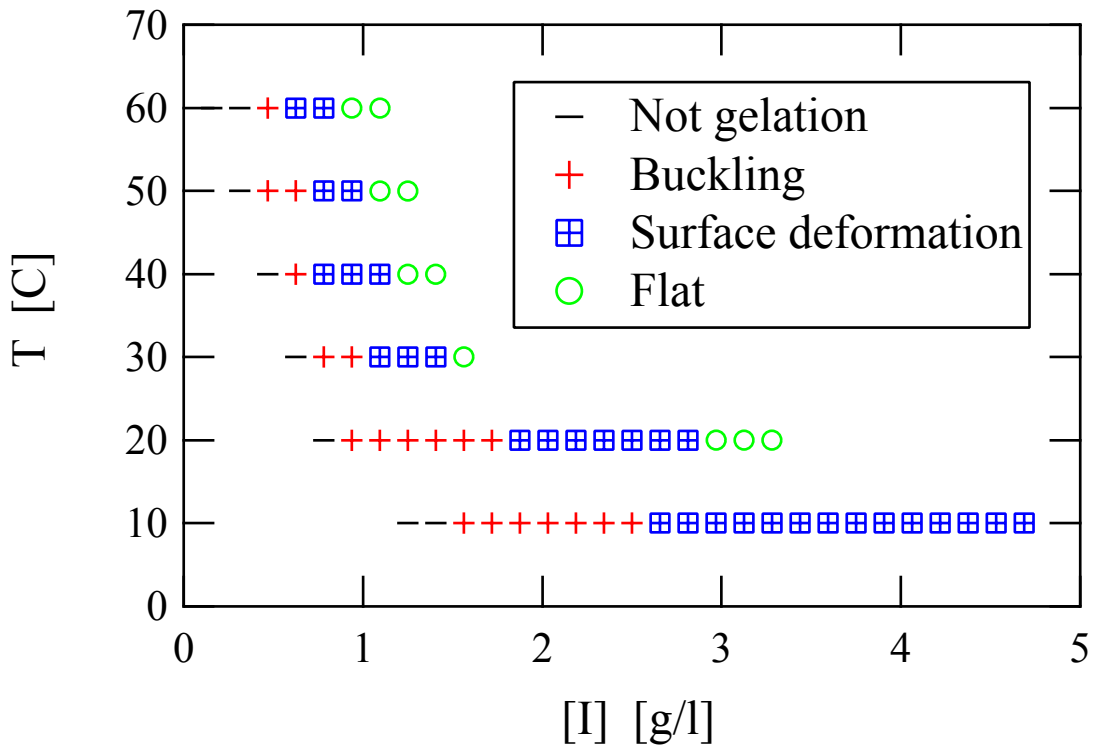

Fig. 2 Phase diagram of DMAA gel slabs. [*I*] is the initiator concentration. In low temperature regime, pattern appearing range is wider than AA gel case. No clear stripe patterns are observed. Other features are similar to the phase diagram of AA gel surface pattern [10].

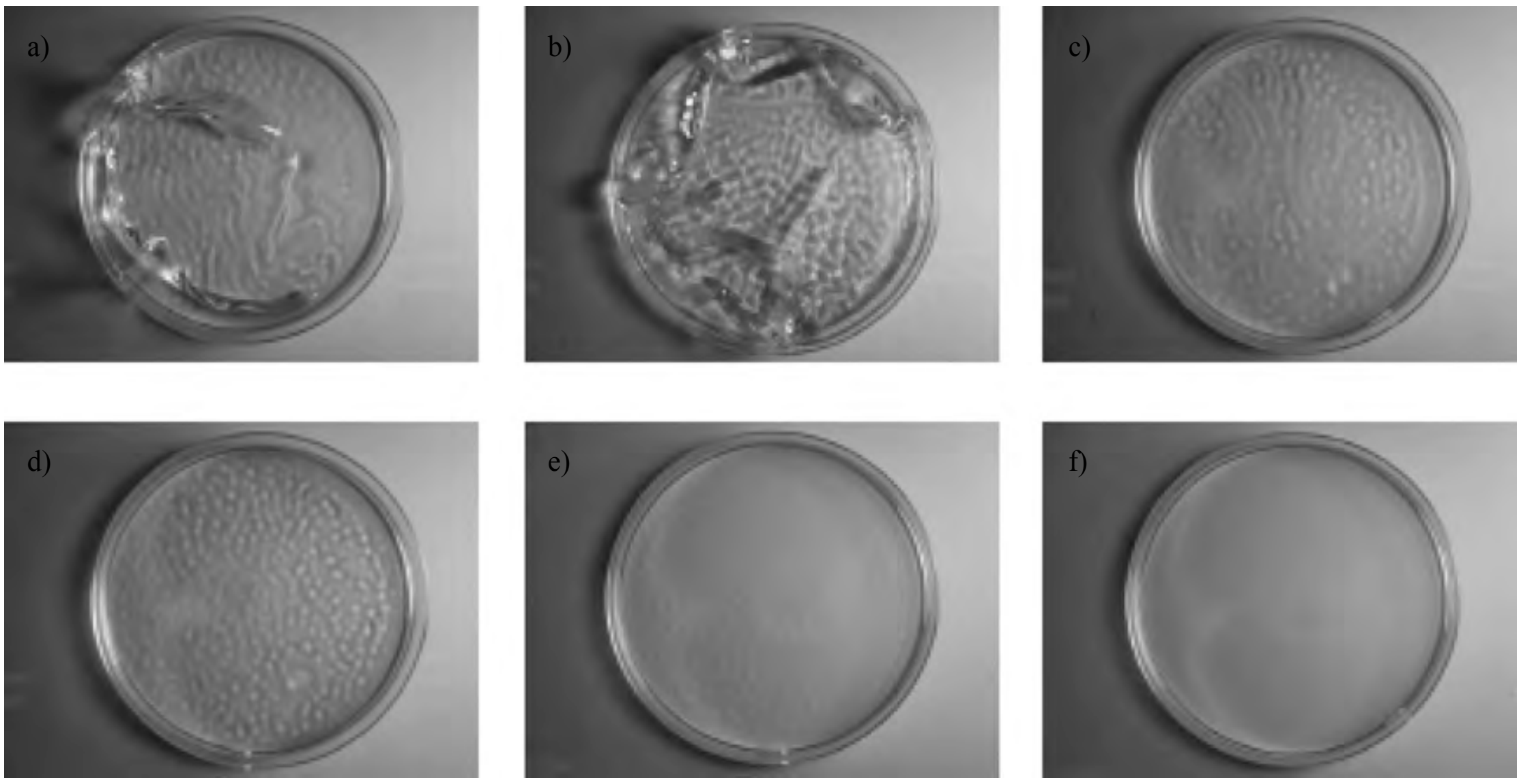

Fig. 3 Examples of DMAA surface deformation and buckling. 1.8 ml DMAA, 4 mg BIS, 70 ml TEMD, and 11 ml deionized water are used. Enviromental temperature is controlled as 30 degree Celsius. The amount of initiator (APS) is varied as **a** 10, **b** 12, **c** 14, **d** 16, **e** 18, **f** 20 mg, respectively.

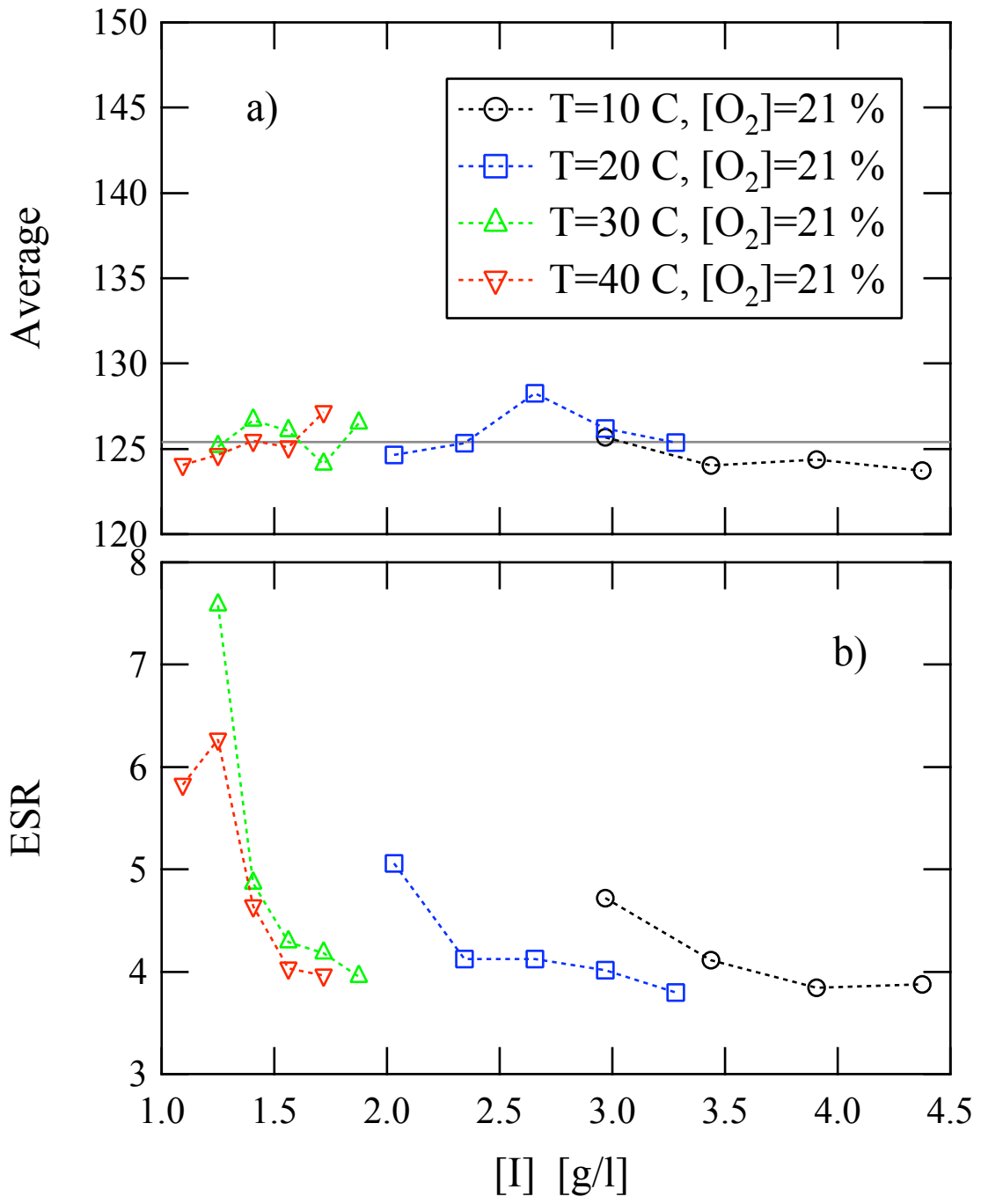

Fig. 4 **a** Average and **b** Effective Surface Roughness (ESR) of varying initiator concentration $[I]$ and temperature $T$. Constant average and varying ESR can be observed as a function of $[I]$.

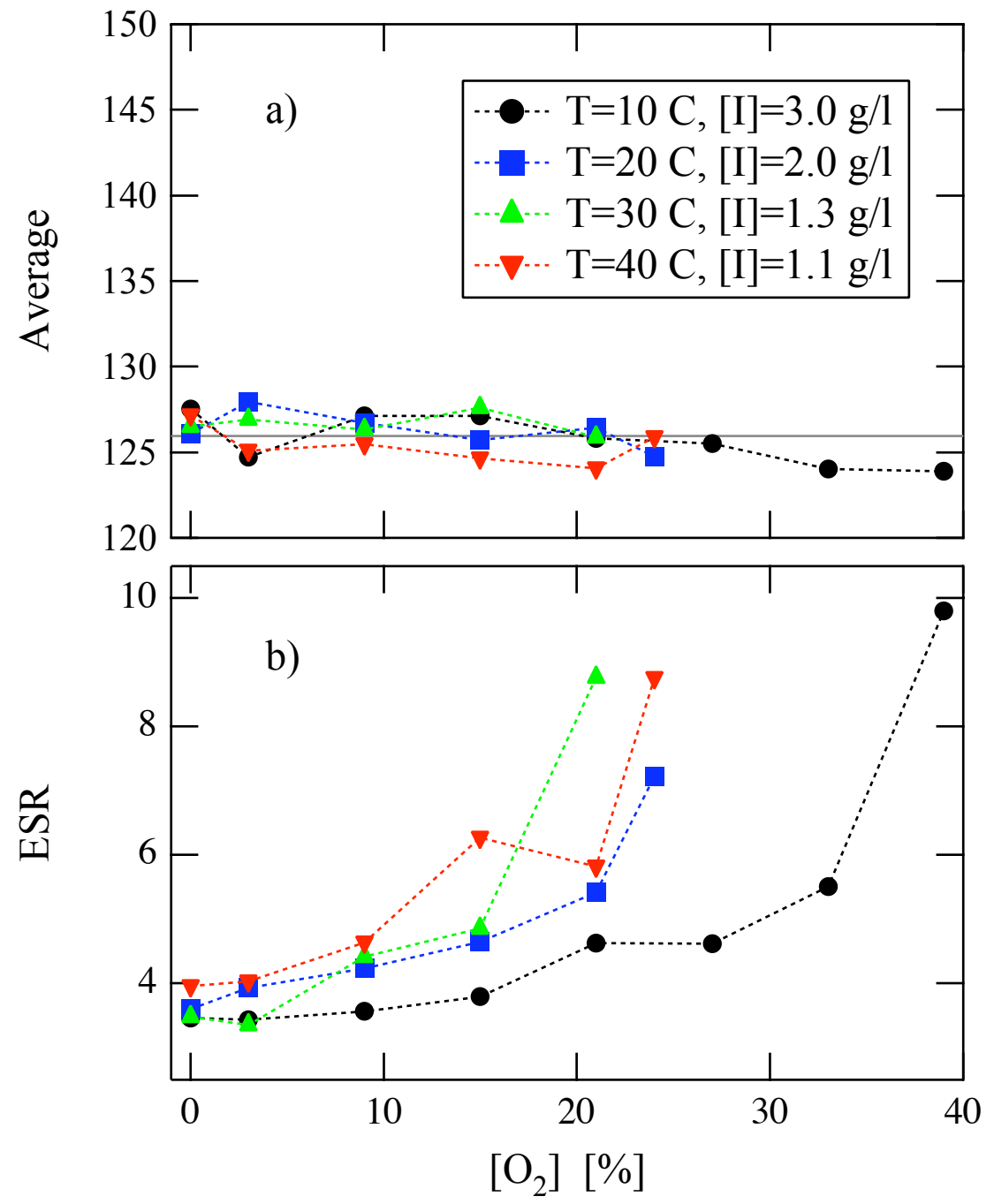

Fig. 5 **a** Average and **b** ESR of varying oxygen concentration and temperature. Where $[O_2]$ is the oxygen concentration. Qualitative behavior of average and ESR is similar to Fig. 4 case, except that the effect of $[I]$ and $[O_2]$ is opposed each other.

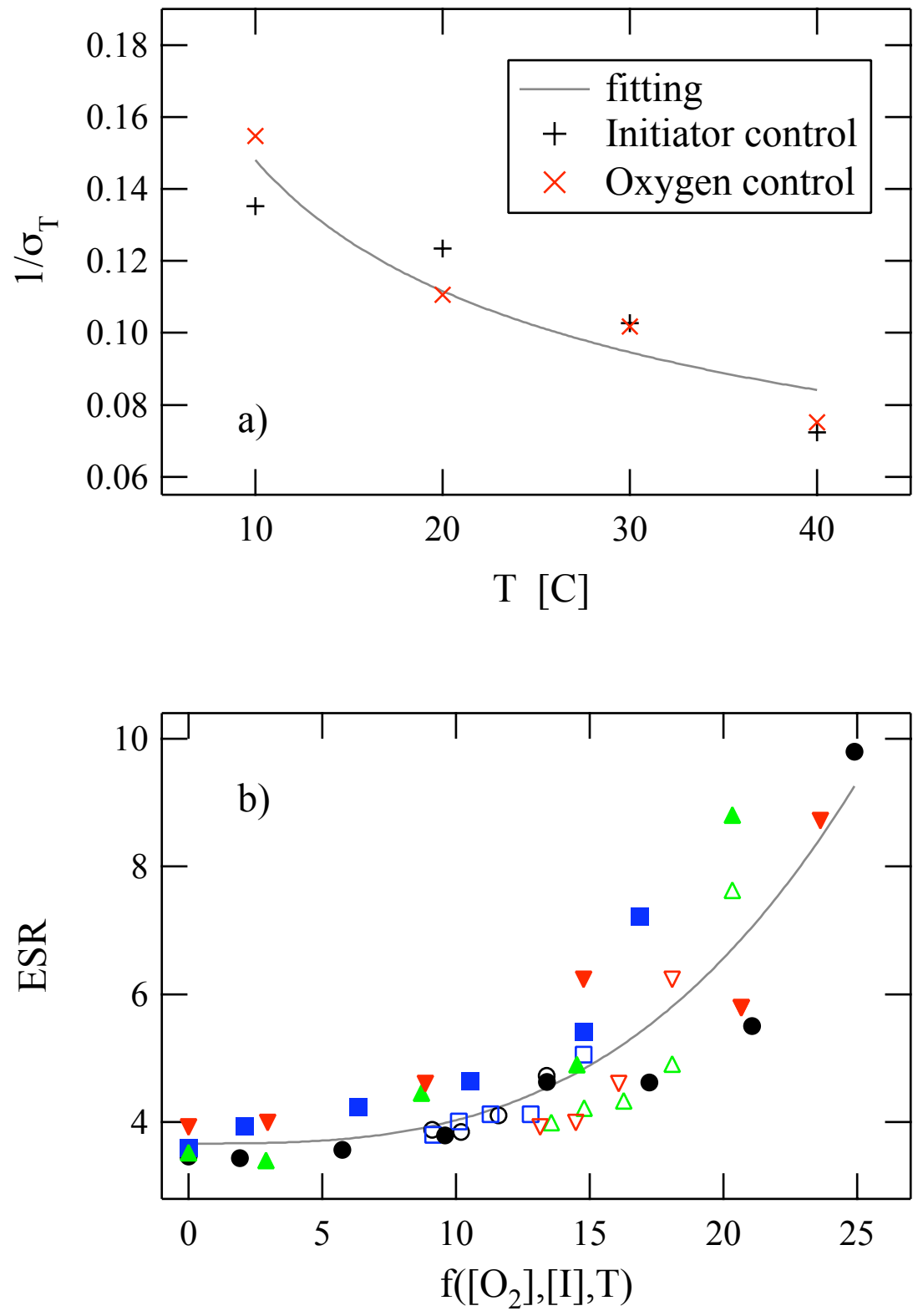

Fig. 6 **a** Scaling of temperature factor and **b** data collapse of ESR by scaling form Eq. (4). All different conditions data are roughly collapsed to the master curve.